\documentclass[intlimits,twoside,a4paper]{article}

\usepackage{amsmath,amssymb}
\usepackage{graphicx}
\usepackage{siunitx}
\usepackage{wrapfig}

\usepackage[T2A]{fontenc}
\usepackage[cp1251]{inputenc}

\usepackage{cmpj2}

\issue{2012}{15}{3}{33401}

\doinumber{10.5488/CMP.15.33401}

\title[ Electro-optical memory of the NLC doped by MWCNTs]%
{Electro-optical memory of a nematic liquid crystal doped by multi-walled carbon nanotubes}
\author[L.~Dolgov \textsl{et al.}]{L.~Dolgov\refaddr{label1}, O.~Yaroshchuk\refaddr{label1}\thanks{E-mail: o.yaroshchuk@gmail.com}\,, S.~Tomylko\refaddr{label1}, N.~Lebovka\refaddr{label2}\thanks{E-mail: lebovka@gmail.com}}
\addresses{
\addr{label1} Institute of Physics, Nat. Acad. of Sci. of Ukraine, 46 Nauki Prospect, 03680 Kyiv, Ukraine
\addr{label2} Institute of Biocolloidal Chemistry named after F.~D. Ovcharenko, Nat. Acad. of Sci. of Ukraine,\\ 42 blvd. Vernadskogo, Kyiv 03142, Ukraine}

\authorcopyright{L.~Dolgov, O.~Yaroshchuk, S.~Tomylko, N.~Lebovka, 2012}

\date{Received February 23, 2012, in final form May 10, 2012}
\begin{document}

\maketitle

\begin{abstract}

A pronounced irreversible electro-optical response
(memory effect) has been recently observed for nematic liquid crystal (LC)
EBBA doped by multi-walled carbon nanotubes (MWCNTs) near
 the percolation threshold of the MWCNTs ($0.02\div 0.05$ wt.~\%).
It is caused by irreversible homeotropic-to-planar reorientation of LC in an electric field. This feature is explained by electro-hydrodynamically stimulated dispergation of MWCNTs in LC and by the formation of a percolation MWCNT network which acts as a spatially distributed surface stabilizing the planar state of the LC. This mechanism is confirmed by the absence of memory in the EBBA/MWCNT composites, whose original structure is fixed by a polymer. The observed effect suggests new operation modes for the memory type and bistable LC devices, as well as a method for \textit{in situ} dispergation of carbon nanotubes in LC cells.


\pacs 42.70.Df, 42.79.Kr, 77.22.Ch, 77.84.Lf, 81.07.De
\end{abstract}

\section{Introduction}

In the recent years, carbon nanotubes (CNTs) cause an ever increasing
interest as a dopant for a LC medium. Owing to an effective
interaction with a LC and extremely high aspect ratio ($100\div 1000$),
CNTs can be well integrated into a LC matrix. The CNTs can be
ordered in the LC host and their alignment can be controlled by
external field due to collective reorientation by the LC
~\cite{Lynch2002,Dierking2004,Scalia2007,Kumar2007}. On the other
hand, CNTs bring a number of improvements for LC layers used in
the electro-optical  devices. Addition of a minute amount of CNTs
 makes it possible to reduce the response time and drive a voltage and thus
suppresses a parasitic backflow and image sticking typical of LC
cells \cite{Lee2004,Rahman2009,Huang2006}.

Recently, studying the LC-CNTs composites based on nematic LC p-ethoxy-benzylidene-p-n-butylaniline (EBBA), we revealed that these composites may demonstrate very unusual memory-type electro-optic response \cite{Dolgov2008,Dolgov2009}. The present paper extends the knowledge of the experimental conditions necessary for this effect to occur and clarifies its nature. It is found that the effect is due to a strong dispersion of carbon nanotubes and the formation of a fine CNT network stabilizing the LC texture formed under the electric field. It is shown that this mechanism can be blocked by fixing the initial structure of CNTs using a photopolymer network.

\section{Materials and methods}

EBBA was used as a liquid crystal because in the preliminary studies the composites based on this LC demonstrated the highest efficiency of electro-optic memory \cite{Dolgov2008}.
The EBBA (Reakhim~Ltd., Russia) with negative dielectric
anisotropy ($\Delta\varepsilon =-0.13$ at $40^{\mathrm{\circ}}$C) was purified by
fractional crystallization from the $n$-hexane solution. After
purification, it exhibited the nematic mesophase between 36 and $79^{\mathrm{\circ}}$C.

The MWCNTs (SpetsMash Ltd., Ukraine) were prepared from ethylene
by chemical vapor deposition method \cite{Melezhyk2005}.
Typically, such MWCNTs have the outer diameter of about $10\div 30$~nm and
the length of about $5\div 10$~\SI{}{\micro\meter}.

The composites were prepared by 20 min ultrasonic stirring of EBBA
/MWCNT mixtures at the temperature of $40^{\mathrm{\circ}}$C, the frequency of 22~kHz and the output power of
150 W. The concentration of MWCNTs, $C$, varied in the range
of $0.004\div 0.5$~wt.~\%. Doping by MWCNTs has not essentially effected the phase transition temperatures of EBBA/MWCNT
composites.

The electro-optical cells were made from glass substrates
containing patterned ITO electrodes and aligning layers of
polyimide AL2021 (JSR, Japan), which produced homeotropic
alignment. The polyimide layers were rubbed by a velvet cloth in
order to provide a uniform planar alignment of LC in the field-on
state. The cells were assembled so that the rubbing directions of
the opposite aligning layers were antiparallel. A cell gap was
maintained by \SI{16}{\micro\meter} glass spacers. The cells were filled with
EBBA/MWCNT composite using the capillary method.

The electro-optical measurements were carried out using the
experimental setup described in \cite{Kovalchuk2001}. The cell was set between two
 crossed polarizers so that the angle between the
 polarizer axes and the rubbing direction was $45^{\mathrm{\circ}}$.
 The sinusoidal voltage of $0\div 60$~V (at the frequency $f=2$~kHz)
 was applied to the cell. The voltage was stepwisely increased from 0 to 60~V and then decreased back to 0~V;
 the total measuring time, i.e., the time of voltage application,
 was about 1 min. The transmittance of the samples was
 calculated as $T=(I_{\mathrm{out}}/I_{\mathrm{in}})\cdot 100$\%, where $I_{\mathrm{in}}$ and $I_{\mathrm{out}}$
 are intensities of the incident and transmitted light,
 respectively. The structure of EBBA/MWCNT composites
 was monitored by observing the filled cells placed
 between crossed polarizers, both with naked eye and by means of  an
 optical polarization microscope.

\section{Results and discussion}
Let us first consider the features of electro-optical response.
Electric field application to the \linebreak EBBA/MWCNT samples resulted in
reorientation of LC from the initial homeotropic to the planar
state. At first, this transition occurred in the middle part of
the cell. With the voltage increase, the area of planar alignment
continuously extended approaching the aligning substrates. The
described reorientation continuously changed the transmittance of the
samples placed between a pair of crossed polarizers.

\begin{figure}[ht]
\centerline{
\includegraphics[width=0.65\textwidth]{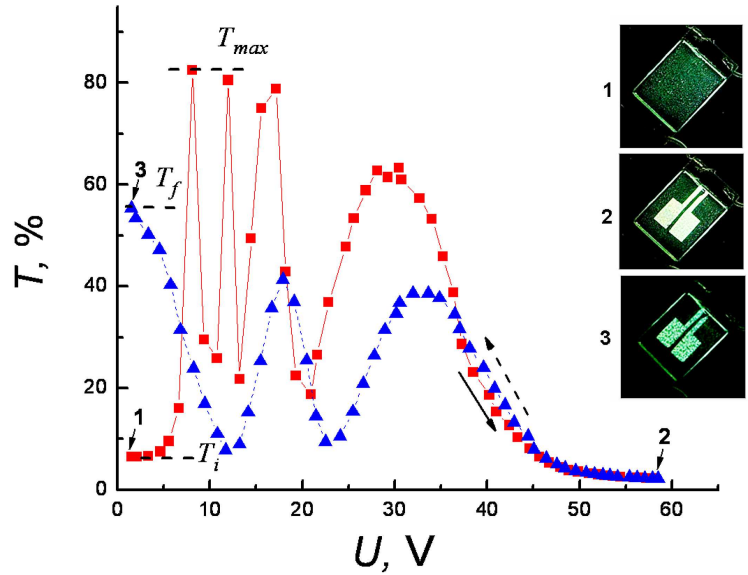}
}
\caption{
(Color online) Optical transmittance $T$ versus the applied voltage $U$ curves for
the cell containing EBBA/MWCNT composite with $C=0.05$ wt.~\%.
 The solid and dotted lines correspond to the voltage $U=60$~V
 ramp up and down, respectively. The $T_{\mathrm{i}}$, $T_{\mathrm{f}}$ and $T_{\mathrm{max}}$ values
 correspond to the initial, final and maximal transmittance,
 respectively. The inset shows photos of the cell viewed
 between the crossed polarizers before voltage application (1),
 at $U=30$~V (2) and in the final state, i.e., after the voltage shutting-off (3).}
 \label{Figure1}
\end{figure}
\begin{figure}[!h]
\centerline{
\includegraphics[width=0.6\textwidth]{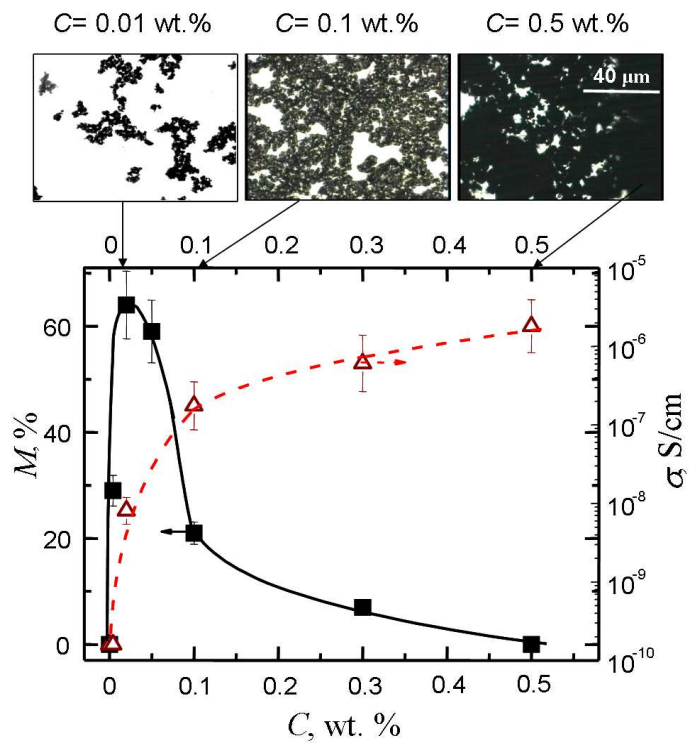}
}
\caption{
(Color online) Electro-optical memory coefficient $M$ and electrical conductivity
$\sigma$  versus concentration of MWNTs, $C$. The micro-photos
taken in the non-polarized light show the aggregates of MWNTs at
different concentrations.}
 \label{Figure2}
\end{figure}

Figure~\ref{Figure1} shows a typical behaviour of the optical
transmittance $T$ as a function of the applied voltage $U$ at voltage
increase and decrease (solid and dotted curve, respectively) for the sample with irreversible response. The oscillations of $T(U)$ curves suggest that the phase difference
between the ordinary and extraordinary waves changes from about 0
to several $\pi$  in the course of homeotropic-to-planar
reorientation \cite{Blinov1993}.

It is also evident from figure~\ref{Figure1} that $T_{\mathrm{f}}$ is
essentially higher than $T_{\mathrm{i}}$, where $T_{\mathrm{i}}$ and $T_{\mathrm{f}}$ are
zero-field transmittances corresponding to the initial and final
measurement stages, respectively. The same can be clearly seen
from the photos of electro-optical cell (configuration between
crossed polarizes) presented in the inset to figure~\ref{Figure1}. The
electrodes in this cell are patterned so that the electric field
is applied only to two rectangular areas in the middle of the
cell. ``Brightening'' of these areas after the field application
cycle (photograph 3) suggests that the cell to a certain
extent memorizes  the optical state formed under the electric field
(photograph 2). In other words, this sample exhibits the effect of
electro-optical memory. It is convenient to define the efficiency of
the electro-optical memory as
\begin{equation}
M=\frac{(T_{\mathrm{f}}-T_{\mathrm{i}})}{(T_{\mathrm{max}}-T_{\mathrm{i}})}\cdot 100\% \,,\label{Eq01}
\end{equation}
where $T_{\mathrm{max}}$ is the maximum transmittance, observed for a
sample under the voltage (figure~\ref{Figure1}). According to this
definition, the memory efficiency $M$ changes within the range of
$0\div 100$~\%. The case $M\approx 0$, i.e., reversible electro-optical
response, was observed for pure LC EBBA and for EBBA/MWCNT
composites with a very small concentration of nanotubes ($C<0.002$
wt.~\%). The electro-optical memory effect became apparent at a
higher concentration of MWCNTs. The memory efficiency $M$ sharply
increases with $C$, goes through maximum at about $C\approx 0.05$
wt.~\% and gradually decreases to zero with a further increase of
MWCNT concentration (figure~\ref{Figure2}).

The $M$ parameter considerably depends on the history of the
system, the regimes of voltage application and the time of electric field
exposure, $t$. The electro-optical memory effect was observed only
at voltages exceeding a certain threshold value ($U_{\mathrm{c}} \approx
15$~V). At smaller voltages, the reversible electro-optical
response ($M\approx 0$) was observed even at a long time of electric
field application ($t>300$~s). At $U>U_{\mathrm{c}}$, the memory parameter
$M$ was an increasing and saturable function of time $t$. The
higher voltage was applied, the faster value of saturation was
reached. A saturation value of $M$ is more or less the same for
different voltages $U>U_{\mathrm{c}}$.

It was possible to remove the induced memory state by applying
a mechanical stress to the cell or by heating it above the temperature of the
isotropic-nematic transition and by subsequent cooling to the nematic
state. Also, the memory state was partially erasable by
applying a relatively high ($U>50$~V) electric field of low
frequency ($f=10\div 50$ Hz).

To clarify the nature of the described memory effect, the
microscopic structures of EBBA and \linebreak EBBA/MWCNT samples were
investigated. The inset of figure~\ref{Figure2} shows that the
increase of MWCNT concentration $C$ led to the structural
transition from isolated aggregates to a continuous network with the
network coherence growing with $C$. The structural changes under
the applied voltage also considerably depended on the
concentration of nanotubes.

In pure EBBA and in the EBBA/MWNNT composites with small
concentration of the nanotubes ($C< 0.002$ wt.~\%), the homeotropic
orientation transformed to the planar one at the threshold voltage
$U\approx 10$~V. The further voltage increase above $U_{\mathrm{c}}\approx
80$~V led to formation of classical electro-hydrodynamic (EHD)
instabilities in the LC host. The typical lamellar flows in the
form of Kapustin-Williams domains \cite{Blinov1993} were observed
for the voltage interval between 80~V and 110~V. At higher voltages, $U=110\div 120$~V, these instabilities  were transformed into  turbulent flows causing the ``boiling'' of LC structure. Note that the samples relaxed to the initial state
after the voltage switch off.

\begin{figure}[ht]
\centerline{
\includegraphics[width=7 cm]{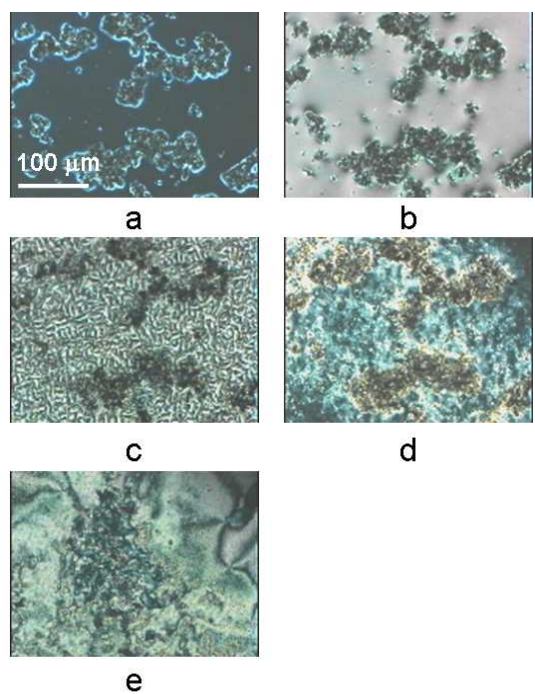}
}
\caption{
(Color online) Microstructures of EBBA/MWNT composite in a sequence of voltage
application steps: (a) before voltage application; (b) under the
voltage of 10~V; (c) under the voltage of 30~V; (d) after exposure
to voltage of 30~V during 60 s; (e) after exposure to voltage of
50~V during 60 s. The concentration of MWNTs in the composite was
$C=0.05$ wt.~\%.}
 \label{Figure3}
\end{figure}

In the EBBA/MWNT composites with $C=0.02\div 0.05$ wt.~\%, electric field initiated the ``boiling'' of the LC phase even at moderate voltages ($U\geqslant  U_{\mathrm{c}}\approx 15$~V). Figure~\ref{Figure3} demonstrates the structural evolution of such
samples under the electric field. The voltage $U=10$~V caused
homeotropic-to-planar reorientation of the LC phase [figure~\ref{Figure3}~(b)]. The subsequent voltage increase above $U_{\mathrm{c}} \approx 15$~V
resulted in the development of EHD flows. These flows arose first in
the vicinity of MWCNT aggregates, presumably, due to a
higher concentration of ionic impurities. The following voltage increase
caused the  intensification of EHD flows and the extension of the areas of
turbulence [figure~\ref{Figure3}~(c)]. The ``boiling'' of LC phase caused
the crushing of large MWCNT aggregates and thus improved the dispergation
of MWCNT phase (EHD-induced dispergation) resulting in the development of EHD flows. Such a behavior has already been noticed for these samples in \cite{Dolgov2008,Dolgov2009,Lisetski2009}. At the first stage of EHD-induced
dispergation, MWCNT aggregates became friable and interactions
between MWCNTs weakened [figure~\ref{Figure3}~(d)]. The longer exposure
to $U>U_{\mathrm{c}}$ and/or increase of voltage amplitude resulted in
further dispersal of these impaired aggregates [figure~\ref{Figure3}~(e)]. The fact that initial bulky aggregates were not
restored after turning off the voltage, suggests that the fine
MWCNT structure was rather stable. The planar LC state realized
under the electric field was conserved too. At first, stable
islands of the planar state arose near the partially destroyed
MWCNT aggregates [figure~\ref{Figure3}~(d)]. With further dispergation
of MWCNT aggregates, new islands of the planar state were formed.
They enlarged, merged and, finally, formed a quasi-continual film
[figure~\ref{Figure3}~(e)]. On the macroscopic level, such non-relaxed
planar structures caused ``brightening'' of electro-optical cells
(figure~\ref{Figure1}), i.e., the memory effect took place.

The EHD instabilities were also observed in highly concentrated
suspensions ($C>0.2$ wt.~\%). Ho\-we\-ver, the intensity of EHD flows and,
thus, the effectiveness of EHD-induced dispergation were essentially
lower than in the composites with $C=0.02\div 0.05$ wt.~\%. Moreover,
the microscopic observations did not reveal any considerable
stabilization of the planar structure after the electric field was
off. This correlates well with a tiny memory efficiency in this
range of MWCNT concentrations (figure~\ref{Figure2}). Such
peculiarities at $C>0.2$ wt.~\% might be caused by an increased
conductivity and, thus, by a decreased actual voltage applied to the
suspension layers.

As noted above, residual transmittance of LC-MWCNT cells was
caused by the random planar alignment memorized after the field
turn-off. The next stage of the present research was aimed at
finding the force that maintains such a state of alignment. The
hint was obtained from the results for LC-aerosil
composites. It is known that such a system possesses a pronounced
electro-optical memory caused by the orientational transition of
LC from spatially random to oriented state
~\cite{Kreuzer1992,Glushchenko1997}. The oriented state was
metastable in the zero field, because it was maintained by the
network of aerosil particles. It was reasonable to assume that a
similar network arises in the composites based on CNTs. Due to
high conductance of CNTs, the existence of their network can be easily
proved by the conductivity measurements.

The conductivity   $\sigma$ vs. MWCNT concentration $C$ curve is
presented in figure~\ref{Figure2}. This growing curve demonstrates a
typical percolation behavior. The initial steep increase
transforms into a moderate increase at $C_{\mathrm{c}} \approx 0.05$ wt.~\%,
which is the percolation threshold of electric conductivity. At
$C\geqslant  C_{\mathrm{c}}$\,, the electric current mainly flows through continuous
network of MWCNTs formed in LC. A moderate grow of $\sigma(C)$
above $C_{\mathrm{c}}$ is caused by embranchment of this network and, thus,
 the formation of new conductivity channels.

To maintain the planar orientation of LC, the MWCNT network should
be sufficiently strong to resist the elastic tensions. In view of
this, the mechanical rigidity percolation should be considered
additionally to the conductivity percolation. Generally, the
rigidity percolation, corresponding to the sol-gel transition, is
characterized by a threshold concentration $C_{\mathrm{m}}$ higher than
$C_{\mathrm{c}}$. According to rough estimations, $C_{\mathrm{m}}\approx 1.6 C_{\mathrm{c}}$
~\cite{Sahimi1998,Moraru2004}.

Assuming that this correlation is valid for the studied system and
that $C_{\mathrm{c}}\approx 0.05$ wt.~\%, one can obtain that $C_{\mathrm{m}}\approx
0.08$ wt.~\%. Note that this value is by an order of magnitude
lower than the value $C_{\mathrm{m}}\approx  1$~wt.~\% estimated for
LC-aerosil suspensions \cite{Kreuzer1992,Glushchenko1997}. This
might be caused by extremely high aspect ratio of MWCNT particles,
capable of forming the  connected structures at such low concentrations.

Finally, to test the proposed model, the following experiment was carried out. The EBBA/MWCNT mixture with $C=0.05$ wt.~\% was doped with a small amount ($C_{\mathrm{p}}=1$ wt.~\%) of pre-polymer composition NOA65 from Norland. After ultrasonic agitation the mixture was filled into the above-described cell and exposed to UV light from a mercury lamp (15 mW/cm$^2$,~3~min) to polymerize the pre-polymer component. This way, we intended to stabilize the initial MWCNT structure and thus block the memory process. According to figure~4, this scenario is indeed implemented: the structure of bulky aggregates of MWCNT is not affected by intense EHD flows. Obviously, the formed polymer sticks the carbon nanotubes together within the aggregates firmly fixing their original structure. Since bulky aggregates are not crushed, a fine MWCNT network does not arise and thus there is no  factor stabilizing LC alignment corresponding to memory. This explains the reversible electro-optic response obtained for this cell (figure~5). Thus, the proposed model enables us to predict the behavior of LC-MWCNT composites under their doping with a polymer.

\begin{figure}[!ht]
\centerline{
\includegraphics[width=7.5cm]{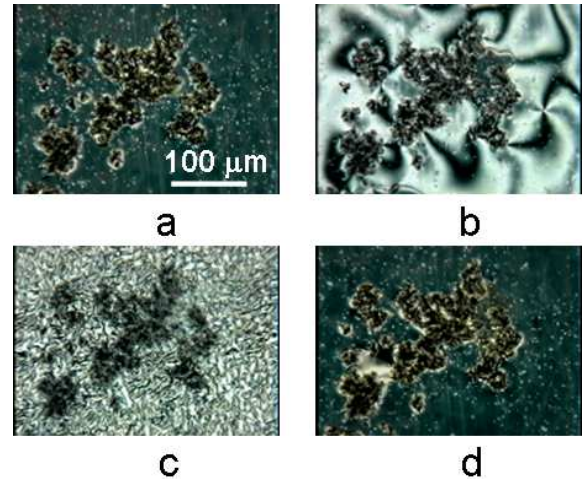}
}
\caption{
(Color online) Microstructures of EBBA/MWCNT/NOA65 ($C=0.05$ wt.~\%, $C_{\mathrm{p}}=0.1$ wt.~\%) composite in the sequence of voltage application steps; (a) before voltage application; (b) under the voltage of 15~V; (c) under the voltage of 35~V; (d) after exposure to voltage 35~V during 60~s.}
 \label{Figure4}
\end{figure}
%
\begin{figure}[!ht]
\centerline{
\includegraphics[width=10cm]{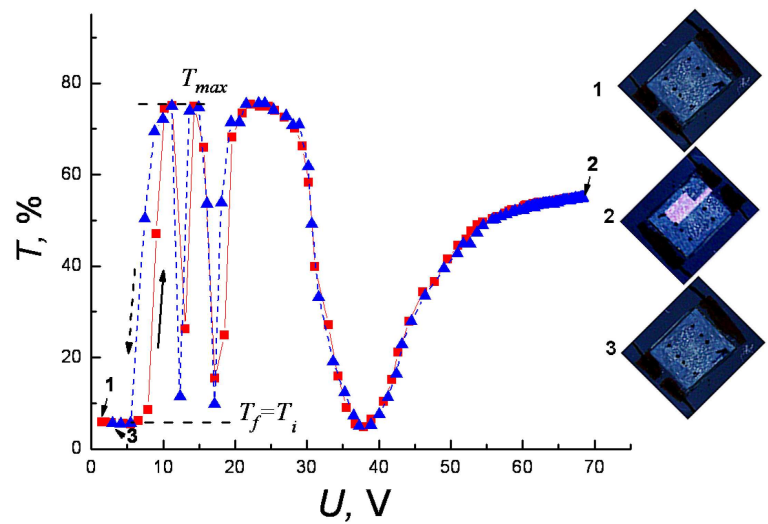}
}
\caption{
(Color online) Optical transmittance $T$ versus the applied voltage U curves
for the cells containing EBBA/MWCNT/NOA65 composite with
$C=0.05$ wt.~\% and $C_{\mathrm{p}}=0.1$ wt.~\%. The solid and
dotted lines correspond to the voltage 35~V ramp up and down,
respectively. The inset shows photos of the cell viewed
between the crossed polarizers before voltage
application (1), at $U=20$~V (2) and in the final off state (3).}
\label{Figure5}
\end{figure}

\section{Conclusions}
\begin{figure}[!ht]
\centerline{
\includegraphics[width=10.5cm]{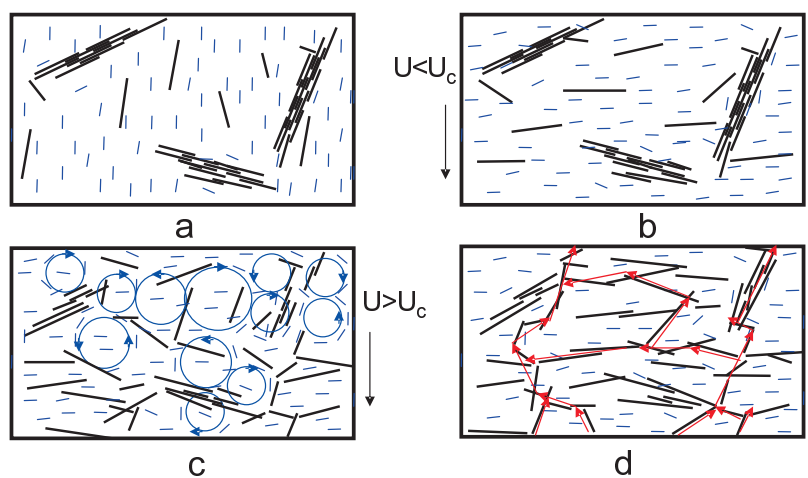}
}
\caption{
(Color online) Schematic presentation of the process of memory state formation.
(a) initial structure of LC-MWCNT samples characterized by bulky
MWCNT aggregates and homeotropic LC alignment; (b)
homeotropic-to-planar reorientation at the voltage above the
Frederiks transition threshold; (c) development of EHD flows above
the critical voltage $U_{\mathrm{c}}$ and crushing of MWCNT aggregates; (d)
formation of fine MWCNT network stabilizing the planar state of
LC. The arrows in (d) show conductivity channels in the MWCNT
network. The arrows near (b) and (c) structures mark the direction of
the applied electric field.}
\label{Figure6}
\end{figure}

In summary, the following mechanism, schematically demonstrated in
figure~\ref{Figure6}, is responsible for the effect of
electro-optical memory. Initially, LC is homeotropically aligned
and MWCNT are well aggregated [figure~\ref{Figure6}~(a)]. Electric
field application leads to LC reorientation from homeotropic to
planar state [figure~\ref{Figure6}~(b)] and to the development of
electro-hydrodynamic flows in the LC phase [figure~\ref{Figure6}~(c)].
These flows crush bulky MWCNT aggregates, thus opening a way for the
formation of a fine MWCNT network, supporting the planar LC
alignment after turning the electric field off  [figure~\ref{Figure6}~(d)]. This mechanism is effective in the limited range
of MWCNT concentrations. On the one hand, $C$ should be higher
than the rigidity percolation threshold $C_{\mathrm{m}}$. On the other hand,
it should not be so high as to prevent a large electric
conductivity and, thus, a low actual voltage applied to suspension
that disables the memory process described above.

This mechanism suggests that the memory effect depends on the interaction of nanotubes with each other and with the LC, and on the strength of homeotropic anchoring at boundary substrates. In addition to the description of LC-MWCNT composites, this mechanism makes it possible to predict the response of such composites when they are doped with a polymer.

The effect of electro-optic memory suggests interesting solutions
for the memory type and bistable LC devices. Besides, the observed
crushing of MWCNT aggregates by EHD flows is quite promising as a
practical method for \textit{in situ} dispergation of MWCNT in a LC
medium.

The work was sponsored by the National Academy of Sciences of
Ukraine (grant No. 10--07--H) and ``Dnipro'' program of
Ukrainian-French scientific cooperation (grant No. M/16-2009)

\ukrainianpart

\title{Електрооптична пам'ять нематичних рідких кристалів допованих багатошаровими вуглецевими нанотрубками}

\author{Л.~Долгов\refaddr{label1}, О.~Ярощук\refaddr{label1}, С.~Томилко\refaddr{label1}, М.~Лебовка\refaddr{label2}}

\addresses{
\addr{label1} Інститут фізики НАНУ, проспект Науки 46, 03680 Київ, Україна
\addr{label2} Інститут біоколоїдної хімії імені Ф.~Д.~Овчаренко, НАНУ, бульвар Вернадського 42, 03142 Київ, Україна
}

\makeukrtitle

\begin{abstract}

\tolerance=3000%
В нематичному рідкому кристалі (РК) ЕББА, допованому
багатошаровими вуглецевими нанотрубками (БШВНТ), поблизу
перколяційної концентрації БШВНТ ($0.02\div 0.05$~ваг.~\%) нещодавно спостережено
виразний необоротний електрооптичний відгук (ефект пам'яті).
Цей ефект викликаний необоротною переорієнтацією РК в електричному полі з гомеотропного у планарний стан. Ця особливість пояснюється електрогідродинамічно стимульованою диспергацією БШВНТ в РК і формуванням перколяційної сітки БШВНТ, яка діє як просторово розподілена поверхня, що стабілізує планарний стан РК. Цей механізм підтверджується відсутністю пам'яті в композитах ЕББА/БШВНТ, у яких початкова структура зафіксована полімером. Цей ефект може бути корисним для розробки нових типів запам'ятовуючих та бістабільних РК пристроїв, а також для підсилення диспергації БШВНТ  безпосередньо в РК комірках.

\keywords вуглецеві нанотрубки, рідкий кристал, ефект пам'яті

\end{abstract}

\end{document}